\documentclass[a4paper,12pt]{article}
\pdfoutput=1
\usepackage{amsmath}
\usepackage{amssymb}
\usepackage{amsfonts,amsthm,dcolumn}
\usepackage{caption}
\usepackage{graphicx,wrapfig}
\usepackage{subfigure}
\usepackage{color}
\usepackage{url}

\RequirePackage[numbers,sort&compress]{natbib}
\RequirePackage{color}
\RequirePackage[colorlinks=true
,urlcolor=blue
,anchorcolor=blue
,citecolor=blue
,filecolor=blue
,linkcolor=blue
,menucolor=blue
,linktocpage=true
,pdfproducer=medialab
,pdfa=true
]{hyperref}

\DeclareGraphicsExtensions{.pdf,.png,.jpg,.jpeg,.eps}

\def\be{\begin{equation}}
\def\ee{\end{equation}\noindent}
\def\ba{\begin{eqnarray}}
\def\ea{\end{eqnarray} \noindent}
\def\nn{\nonumber}

\def\ham{\mathcal{H}}
\def\mce{\mathcal{E}}

\begin{document}
\begin{titlepage}
\thispagestyle{empty}
\begin{flushright}
ICTS/2013/17
\end{flushright}
\bigskip
\begin{center}
\noindent{\Large \textbf{Confining Backgrounds and Quantum Chaos in Holography}}\\
\vspace{2cm} \noindent{
{\Large Pallab Basu\footnote{email:\href{mailto:pallab.basu@icts.res.in}{pallab.basu@icts.res.in}},
Archisman Ghosh\footnote{email:\href{mailto:archisman.ghosh@icts.res.in}{archisman.ghosh@icts.res.in}}}}\\
\vspace{1cm}
{\it 
% $^1$
Department of Physics and Astronomy, \\ University of Kentucky, Lexington, KY 40506, USA \\ \vspace{.5cm} 
% $^2$
International Centre for Theoretical Sciences, \\ Tata Institute of Fundamental Research, \\ Bangalore 560012, INDIA \\}
\end{center}
\vspace{0.3cm}
\begin{abstract}
Classical world-sheet string theory has recently been shown to be nonintegrable and chaotic in various confining string theory backgrounds -- the $AdS$ soliton background in particular. In this paper we study a minisuperspace quantization of the theory and look at properties of the spectrum like the distribution of level spacing, which are indicative of quantum order or chaos. In the quantum spectrum we find a gradual transition from chaotic (Wigner GOE) to integrable (Poisson) regime as we look at higher energies. This is expected since our system is integrable asymptotically, and at higher energies, the dynamics is entirely dominated by the kinetic terms. 
\end{abstract}
\end{titlepage}

\newpage
\section{Introduction}

Understanding string spectra in holographic backgrounds is an intriguing issue. Via gauge gravity duality \cite{Maldacena:1997re,Gubser:1998bc,Witten:1998qj,Aharony:1999ti}, the string spectrum gets mapped to a glueball spectrum in the strongly coupled gauge theory. Many known extensions of the original holographic principle involve a QCD-like confining gauge theory model \cite{Klebanov:2000hb,Maldacena:2000yy}. The goal here is to understand the nature of the QCD spectrum using a dual string background. However the exact string theory spectrum on these simple holographic backgrounds is not completely worked out yet. Even in the simplest case of pure $AdS$, the string spectrum is not fully understood \cite{Beisert:2010jr}. Things are much more intractable for confining backgrounds. It has been shown that classical string motion is nonintegrable in various confining backgrounds \cite{Basu:2011dg,Basu:2011di,Basu:2011fw,Basu:2012ae}\footnote{Chaos as a signature of nonintegrability in string theory was first 
used in \cite{Zayas:2010fs}.}. It seems reasonable to conclude that an analytic understanding is difficult 
to achieve. 

However some statistical properties of the energy spectrum can be understandable. Random Matrix Theory was proposed by Wigner back in 1957 \cite{Wigner:1958} to study hadron spectra in a statistical sense. Later it was understood that appearance of the results of Random Matrix Theory in QCD spectra came under the generic framework of Quantum Chaos. Quantum dynamics of systems which display classical chaos was found to be quite generic and the distribution of level spacing was found to agree with that obtained from a Gaussian Orthogonal Ensemble (GOE) \cite{Berry:1981,Bohigas:1989rq}. Level spacing distribution of eigenvalues for integrable systems on the other had already been shown to obey a Poisson statistics \cite{BerryTabor1977}. However some of the features here are neither universal nor completely understood -- a few of these complications are discussed in \cite{Gutzwiller,Stockmann}.

In this paper, we will try to understand the problem of quantum chaos in an $AdS$ soliton background \cite{Horowitz:1998ha}. In \cite{Basu:2011dg}, the classical dynamics in this background was found to be nonintegrable and chaotic. Here we will try to find the quantum spectrum  in the framework of minisuperspace quantization. It should be noted that our background is asymptotically $AdS$ which is an integrable background \cite{Mandal:2002fs,Bena:2003wd}. Hence it is not clear whether the asymptotic level spacing should really match with that of a Wigner GOE. In fact we will find a gradual transition from Wigner GOE to Poisson distribution for higher energies. 

Quantum chaos in the context of holographic systems has been attempted in \cite{archisman_thesis,PandoZayas:2012ig}. Minisuperspace quantization is implemented in \cite{PandoZayas:2012ig}. However, the authors there do not have a correct normalization of the level-spacing that we have here. Correct normalization here leads us to a Wigner GOE with no free parameters and a $\Delta_3$ statistic for the system. We also discuss the implications of having an asymptotic $AdS$ geometry, and discuss the transition of the system from a chaotic to an integrable regime. Techniques similar to what we have used here, have been applied in a slightly different holographic context in \cite{Saremi:2012ji}.

The rest of this paper is organized as follows. In Section~\ref{sec:setup}, we set our system up and using minisuperspace quantization reduce our problem to that of finding the spectrum for the motion of a particle in a simple potential. In Section~\ref{sec:discussion}, we discuss the properties of the spectrum that we can expect in an integrable and a chaotic quantum system. We end with our results for the $AdS$ soliton background in Section~\ref{sec:results}.

\section{Setup}
\label{sec:setup}
We will work with the $AdS$ soliton background \cite{Horowitz:1998ha} described by the metric\footnote{The metric, embedding and classical equations of motions have been obtained and discussed extensively in \cite{Basu:2011dg}. We will briefly go over the important details, trying to keep the discussion here self-contained.}: 
\begin{align}
& ds^2=\frac{4L^2\alpha'}{d^2}\left\{e^{2ax^2}(-dt^2+T(x) d\theta^2+dw_i^2)+\frac{d^2a^2x^2}{T(x)} dx^2\,\right\},\nn \\
{\rm where\ } & \ T(x)=1-e^{-dax^2}\,,\ a=d/4\,.
\label{eq:adssol_metric1}
\end{align}
The geometry is asymptotically $AdS_{d+1}$ and it caps off in the ``radial'' direction, in these coordinates at $x=0$, where the size $T(x)$ of the compact $\theta$-cycle shrinks to zero. This smoothly cuts off the IR region of $AdS$, dynamically generating a mass scale in the theory, very much like in real QCD -- the dual theory is confining and has a mass gap. The $\theta$-cycle gives a Scherk-Schwarz compactification in the dual theory. The $w$ directions remain non-compact and serve as space directions of the field theory CFT$_{d-1}$.

We will work with closed strings in the geometry, described by the embedding \cite{Basu:2011dg}:
\begin{align}
& t=t(\tau),\ \theta = \theta(\tau),\ x=x(\tau), \nn \\
& w_1=R(\tau)\cos\left(\phi(\sigma)\right),\ w_2=R(\tau)\sin\left(\phi(\sigma)\right)\ {\rm with\ } \phi(\sigma) = \alpha \sigma\,.
\label{eq:embedding}
\end{align}
The string is located at a certain value of $x$ and is wrapped around a pair of $w$-directions as a circle of radius $R$. It is allowed to move along the potential in $x$ direction and change its radius $R$. Here $\alpha\in{\mathbb Z}$ is the winding number of the string.

Upon substitution in the Polyakov action, one obtains the effective Lagrangian for the motion:
\begin{align}
{\cal L}& \propto
\frac{1}{2}e^{2ax^2}\left\{-\dot{t}^2+T(x)\dot{\theta}^2+\dot{R}^2-\alpha^2R^2\right\} + \frac{d^2a^2x^2}{2T(x)} \dot x^2\,,
\label{eq:lag_adssol}
\end{align}
where the dot denotes a derivative w.r.t $\tau$.  
The coordinates $t$ and $\theta$ turn out to be {\em ignorable} and the corresponding momenta $E$ are $k$ constants of motion: 
\begin{align}
p_t = -e^{2ax^2}\dot{t} \equiv E\,, \qquad
p_{\theta} = e^{2ax^2}T(x)\dot{\theta} \equiv k\,. 
\label{eq:Ek}
\end{align}
$R$ and $x$ survive as free coordinates.
The momenta corresponding to these coordinates are:
\begin{align}
p_R =e^{2ax^2}\dot{R}\,, \qquad
p_x =\frac{d^2a^2x^2}{T(x)}\dot{x}\,.
\label{eq:pxpR}
\end{align}
With this one can construct the effective Hamiltonian:
\begin{equation}
\label{eq:ham_adssol}
{\cal{H}} =\frac{1}{2} \left\{
\left(-E^2+\frac{k^2}{T(x)}+p_R^2\right)e^{-2ax^2}
+\frac{T(x)p_x^2}{d^2a^2x^2}
+\alpha^2R^2e^{2ax^2}
\right\}\,.
\end{equation}
The Virasoro constraint equations give $\ham=0$\footnote{The other independent Virasoro constraint is automatically satisfied for our embedding.}.

In \cite{Basu:2011dg} we obtained the equations of motion coming from the Hamiltonian (\ref{eq:ham_adssol}) and solved them classically. Here we are going to do a minisuperspace quantization of the Hamiltonian to find its quantum spectrum.

\subsection{Minisuperspace quantization}\label{sec:minisup}
\begin{figure}[h!]
\centering
\subfigure[$k=4$.]{
\includegraphics[scale=0.8]{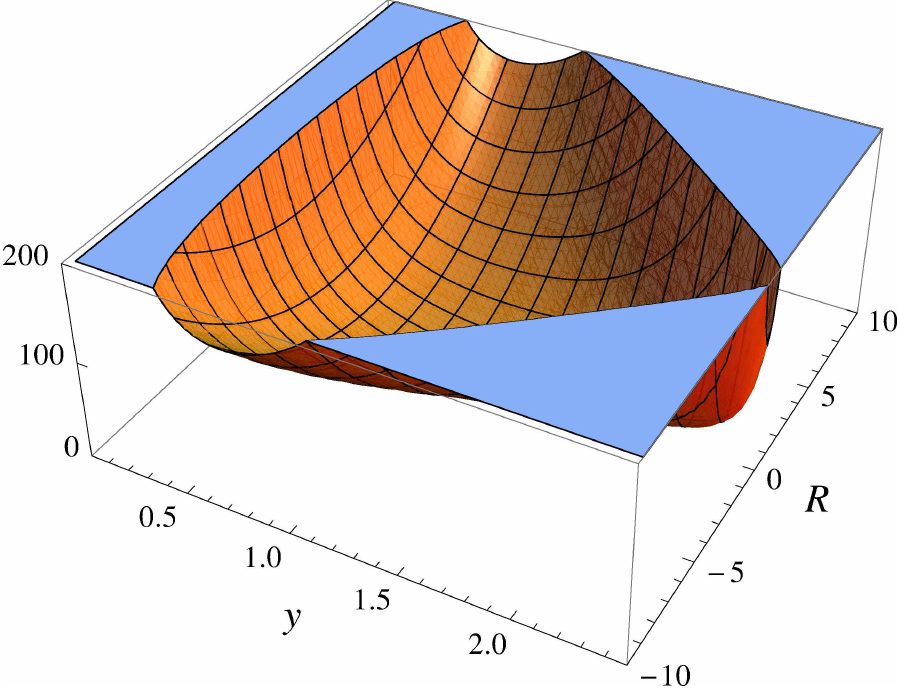}
\label{fig:pot_k4}
}
\subfigure[$k=9$.]{
\includegraphics[scale=0.8]{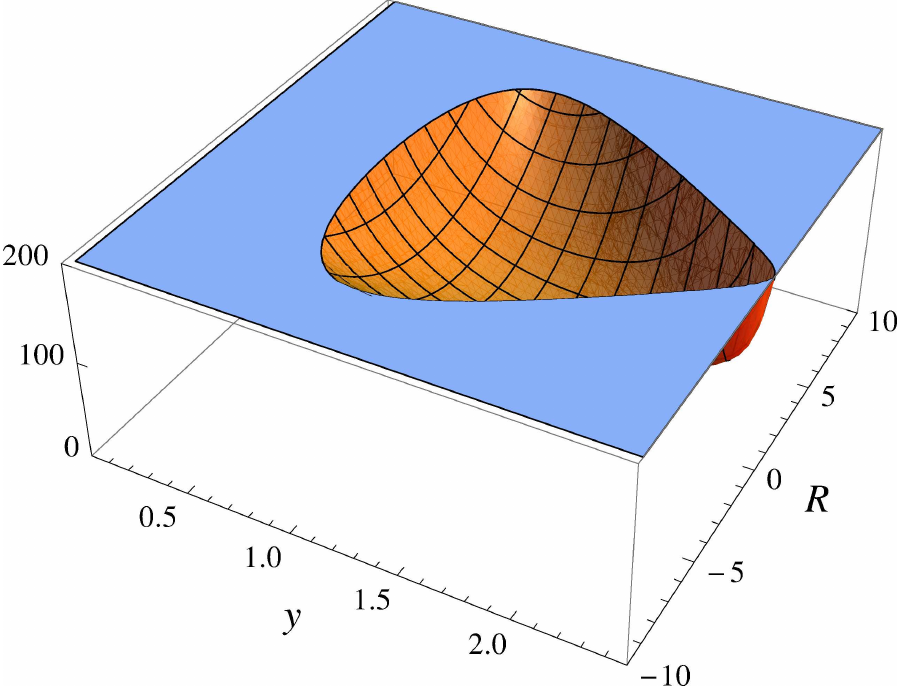}
\label{fig:pot_k9}
}
\caption{Effective potential $\tilde{V}_{\rm eff}(y,R)$ for two values of $k$. We will be interested in $k$ between these two values. On the boundary of the domain $\tilde{V}_{\rm eff}(y,R)\gtrsim200$. So it is consistent to replace the boundary by a hard wall as long as we are interested in eigenvalues $E^2\lesssim200$.}
\label{fig:pot}
\end{figure}

The minisuperspace prescription requires the following substitution in the Hamiltonian:
\begin{align}
\label{eq:minisup-sub}
p_R^2\to-\nabla_R^2\,,\qquad p_x^2\to-\nabla_x^2\,.
\end{align}
Here the Laplacian is calculated w.r.t. the effective metric seen in the Lagrangian (\ref{eq:lag_adssol}):
\begin{align}
\label{eq:eff-metric}
-g_{tt}=g_{RR}=e^{2ax^2}\,,\quad g_{\theta\theta}=e^{2ax^2}T(x)\,,\quad g_{xx}=\frac{d^2a^2x^2}{T(x)}\,.
\end{align}
This gives us the minisuperspace Hamiltonian:
\begin{align}
\label{eq:adssol-minisup}
\ham&=\frac{1}{2}\left\{\left(-E^2+\frac{k^2}{T(x)}-\partial_R^2\right)e^{-2ax^2}-\frac{T(x)}{d^2a^2x^2}\partial_x^2\right. \nn \\ &-\left.\left(\frac{T'(x)}{T(x)}-\frac{1}{x}+6ax\right)\frac{T(x)}{d^2a^2x^2}\partial_x+\alpha^2R^2e^{2ax^2}\right\}\,.
\end{align}
We need to find the eigenvalues of $\ham\psi=0$. The eigenvalue equation takes the form:
\begin{align}
\label{eq:adssol-effham-xR}
E^2\psi(x,R)&=%\frac{k^2}{T(x)}\psi(x,R)
-\partial_R^2\psi(x,R)-f(x)\partial_x^2\psi(x,R)
-g(x)\partial_x\psi(x,R)+V_{\rm eff}(x,R)\psi(x,R)\,.
\end{align}
Here we have defined:
\begin{align}
\label{eq:adssol-effham-fgV}
f(x)\equiv\frac{T(x)e^{2ax^2}}{d^2a^2x^2},\quad & g(x)\equiv\left(\frac{T'(x)}{T(x)}-\frac{1}{x}+6ax\right)\frac{T(x)e^{2ax^2}}{d^2a^2x^2},\nn \\ 
V_{\rm eff}(x,R)&\equiv\frac{k^2}{T(x)}+\alpha^2R^2e^{4ax^2}\,.
\end{align}
With a coordinate transformation $dy=dx/\sqrt{f}$ and a field redefinition
\begin{equation}
\label{eq:adssol-transf1}
\tilde{\psi}=e^\beta\psi\quad{\rm such\ that}\quad\partial_y\beta=\frac{f'-2g}{4\sqrt{f}}\,,
\end{equation}
we get (where prime denotes a derivative w.r.t $x$),
\begin{equation}
\label{eq:adssol-transf2}
f\partial_x^2\psi+g\partial_x\psi=\partial_y^2\psi-\frac{f'-2g}{2\sqrt{f}}\partial_y\psi
=e^\beta\left\{\partial_y^2\tilde{\psi}+\left[\partial_y^2\beta - ( \partial_y \beta )^2\right]\tilde{\psi}\right\}\,.
\end{equation}
The eigenvalue equation now simplifies to:
\begin{equation}
\label{eq:adssol-effham-yR}
E^2\tilde\psi(y,R)= % \frac{k^2}{\tilde{T}(y)}
-\partial_R^2\tilde\psi(y,R)-\partial_y^2\tilde\psi(y,R)+\tilde{V}_{\rm eff}(y,R)\tilde\psi(y,R)\,,
\end{equation}
where,
\begin{eqnarray}
\label{eq:adssol-effham-transf}
&& % \tilde{T}(y)\equiv{}T(x(y)),\quad
\tilde{V}_{\rm eff}(y,R)\equiv\frac{k^2}{T(x(y))}+V(x(y),R)-\left[\sqrt{f} ( \partial_y \beta )' - ( \partial_y \beta )^2 \right] \,.
% \nn \\ {\rm and}&&\tilde\psi(y,R)=f(x(y))\tilde\psi(x(y),R)\,.
\end{eqnarray}
The domain of the problem $x\in(0,\infty)$ is mapped to $y\in(0,y_\infty)$ with $y_\infty\approx2.62$.
The effective potential $\tilde{V}_{\rm eff}(y,R)$, plotted for two values of $k$, is shown in Figure \ref{fig:pot}. Here and henceforth, we choose $d=4$ and $\alpha=1$.

\section{Discussion}
\label{sec:discussion}
\subsection{Quantum chaos in nonintegrable systems}\label{sec:qc}
Quantum chaos\footnote{For good books on the subject we refer the reader to \cite{Gutzwiller,Stockmann}.} is the study of quantum properties of a system whose classical limit is chaotic. A classical description of chaos is usually in terms of trajectories and phase space \cite{Hilborn, Ott}. This framework breaks down in quantum mechanics and a na\"ive extension from the definition of classical chaos is not possible even in principle. The quantum description of a time-independent system is in terms of the energy levels, and one has to look at the spectrum to see if there can be any characteristic properties indicative of chaos.

An interesting property to study turns out to be the distribution of the spacing between adjacent energy levels of the system, normalized such that there is one level per unit interval on an average. For a quantum system whose Hamiltonian is classically integrable, it was shown by Berry and Tabor (1977) \cite{BerryTabor1977} that this level-spacing distribution is quite universally same as that of the spacing between a sequence of random uncorrelated levels, which is a Poisson distribution:
\begin{align}
\label{eq:poisson}
P(s)\simeq\exp(-s)\,.
\end{align}
The next question to ask is whether there is any such universal distribution for systems which are classically chaotic. The models most frequently studied in this context are Billiard systems of Sinai, which come up frequently while talking about classically chaotic Hamiltonian systems as well. It was shown by Berry (1981) \cite{Berry:1981} that the in the quantum spectrum of these systems small differences of eigenvalues are avoided.
It was further demonstrated by Bohigas, Giannoni and Schmit (1984) \cite{Bohigas:1983er} that the level spacing distribution of eigenvalues calculated numerically agree with a good approximation to that of a Gaussian orthogonal ensemble (GOE) of random matrices, first discussed by Wigner (1958) \cite{Wigner:1958}. 
\begin{align}
\label{eq:GOE}
P(s)\simeq\frac{\pi s}{2}\exp{\left(-\frac{\pi s^2}{4}\right)}\,.
\end{align}
\begin{figure}
\centering
\subfigure[Integrable potential]{
\includegraphics[scale=0.38]{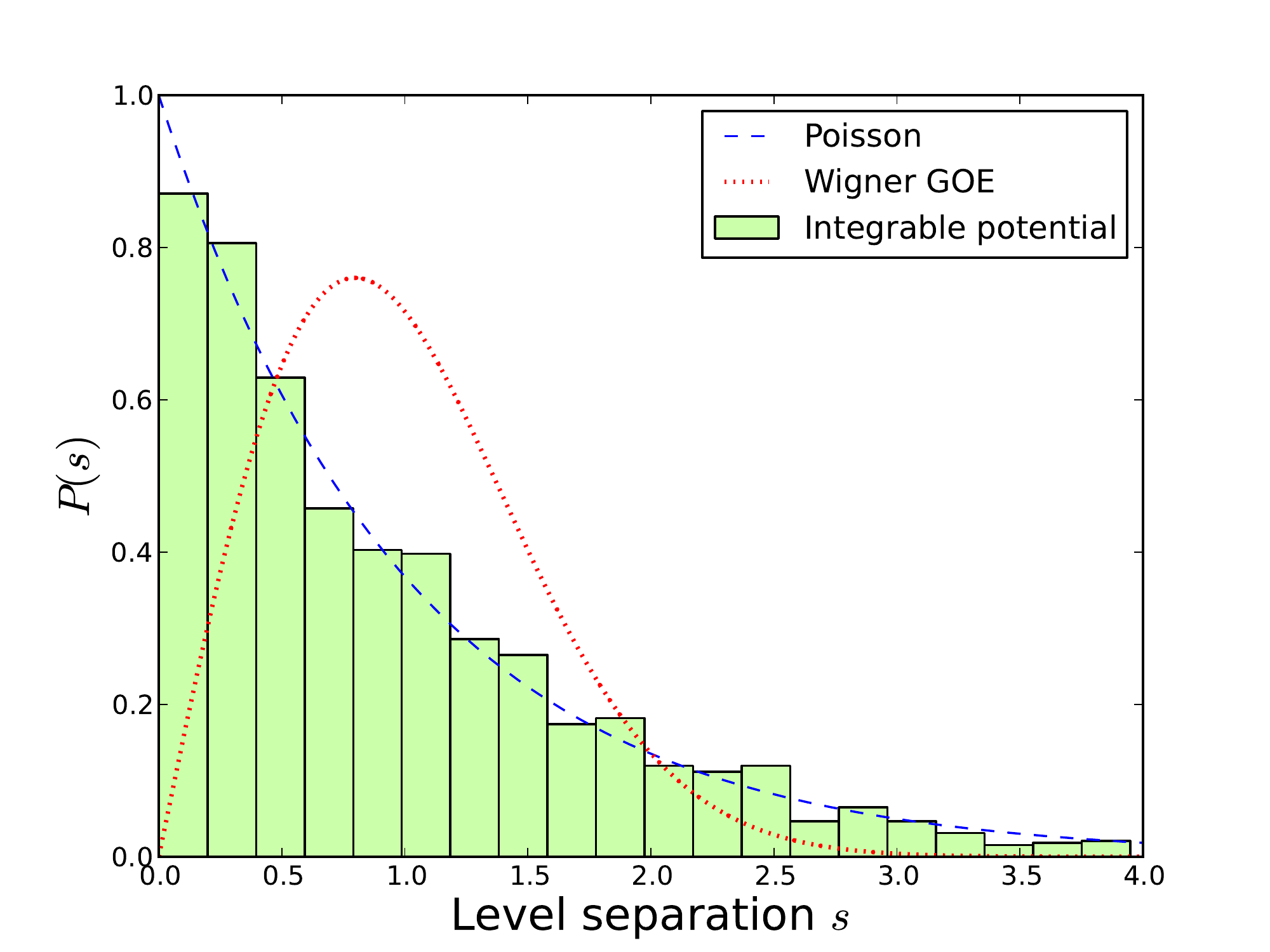}
\put(-90,60){\includegraphics[scale=0.2]{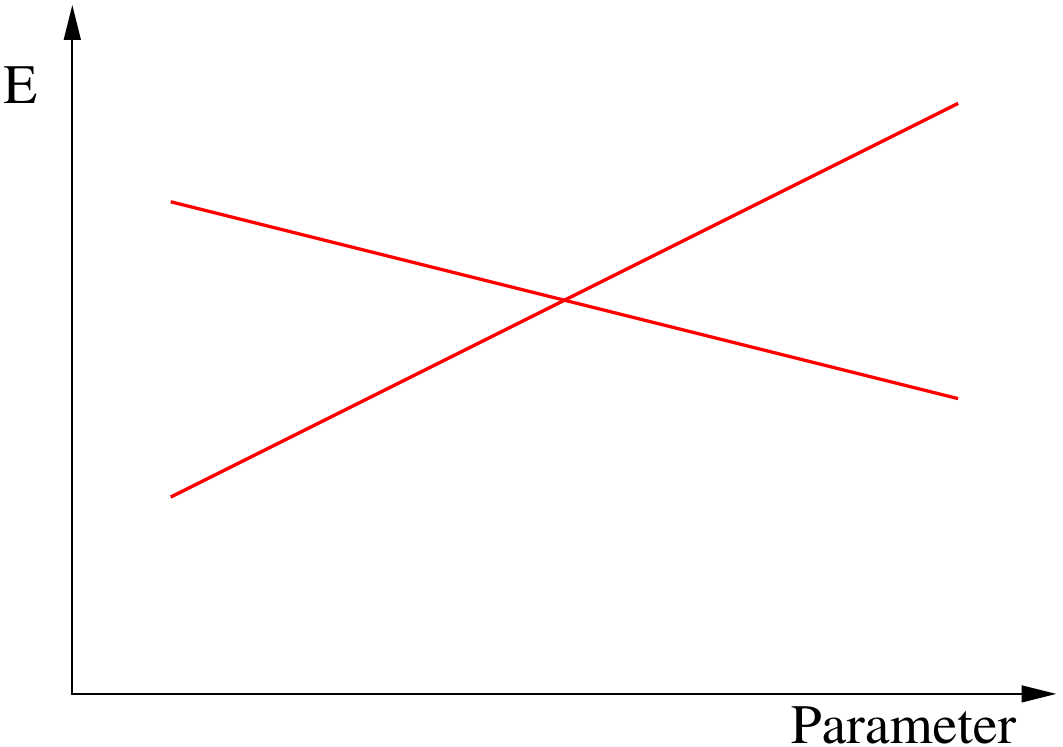}}
\label{fig:qc_box_poisson}
}
\subfigure[Nonintegrable potential]{
\includegraphics[scale=0.38]{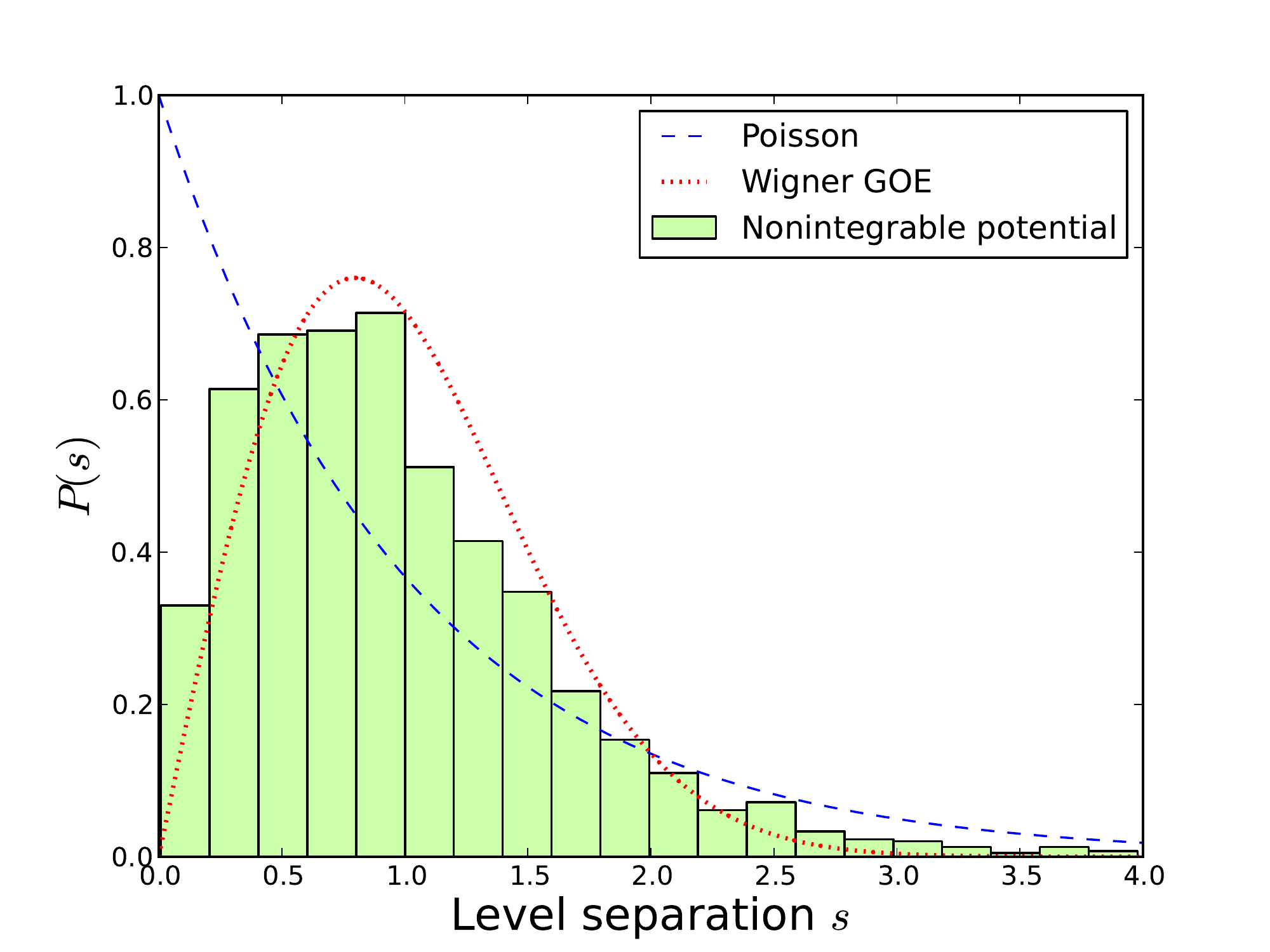}
\put(-90,60){\includegraphics[scale=0.2]{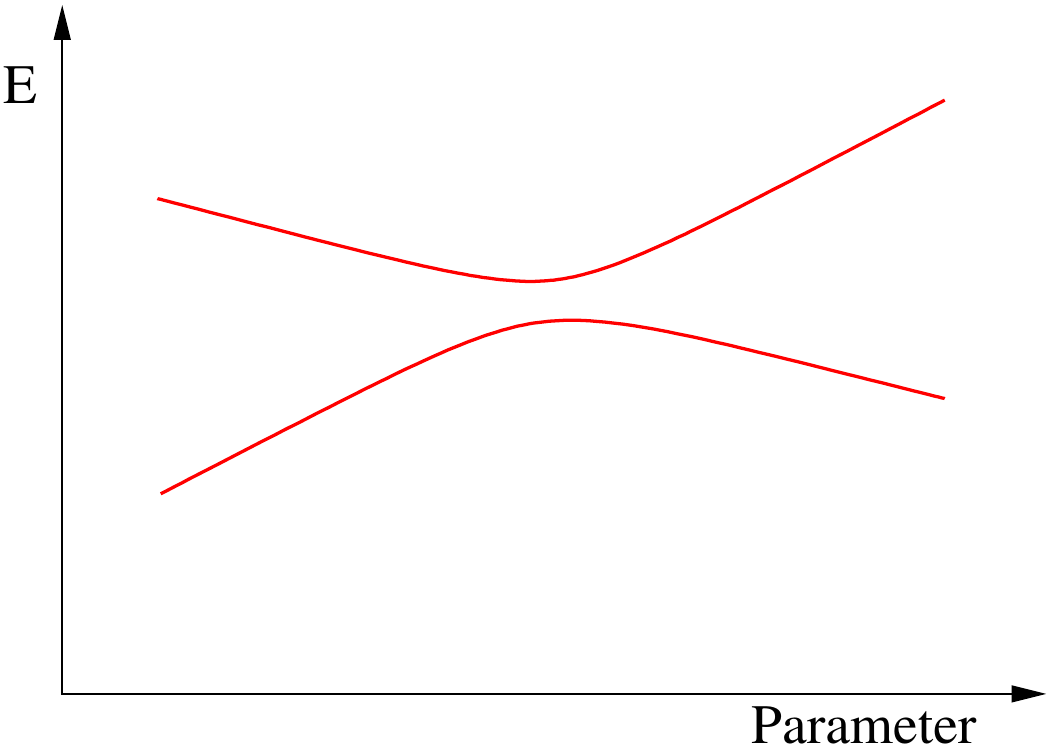}}
\label{fig:qc_box_wigner}
}
\caption{Distribution of level spacing in the spectrum of an integrable and a nonintegrable potential respectively showing agreement with Poisson and Wigner GOE distribution. A particle in a rectangular box of dimensions $\sqrt[4]{e}\times1$ is considered and in the nonintegrable case is deformed by a potential $V(x,y)=\exp\left(\gamma(x-y-1)\right)$ with $\gamma=33.8$. The inset shows how two energy levels vary as a generic parameter is changed. In an integrable system the levels can cross and have degeneracies. In a nonintegrable system the levels repel and hence small differences between energies are avoided.
% Eigenvalues have been calculated using pseudospectral method on a $64\times64$ Tchebychef grid.
}
\label{fig:qc_box}
\end{figure}
A plot of the two distributions is shown in Figure \ref{fig:qc_box}, along with histograms for the example of a particle in a two dimensional box, without and with an additional nonintegrable potential term. The principal qualitative difference here is that $P(s)$ has a maximum at $s=0$ in the Poisson distribution for the integrable case but it goes to zero as $s\to0$ in the Wigner GOE distribution for the chaotic case.

It might initially seem a bit counterintuitive that the level spacing distribution for an integrable system is same as that of randomly spaced levels and that for a nonintegrable system has some additional structure. However one should note that this structure comes from quantum mechanical {\em level repulsion}, which is able to manifest only in the nonintegrable case. Because of level repulsion, small differences in energies are suppressed. In the integrable case, the eigenvalues coming from the different separable sectors are independent of each other. Hence there is no such repulsion. This is systematically demonstrated for two crossing levels in the inset of Figure \ref{fig:qc_box}.

% \begin{figure}
% \centering
% \subfigure[Integrable system]{
% \includegraphics[scale=0.6]{axes_int.pdf}
% \label{fig:axes_int}
% }
% \subfigure[Nonintegrable system]{
% \includegraphics[scale=0.6]{axes_nonint.pdf}
% \label{fig:axes_nonint}
% }
% \caption[Level-repulsion in a quantum chaotic system]{Level repulsion in a nonintegrable quantum system. In an integrable system the eigenvalues can cross and have degeneracies. In a nonintegrable system the eigenvalues repel. Hence small differences between eigenvalues are avoided.}
% \label{fig:axes}
% \end{figure}

The normalized spacing between adjacent levels for a harmonic oscillator is fixed at unity -- the spectrum is a {\em rigid} spectrum. The departure from equal spacing is another characteristic feature of the spectrum distinguishing integrable and nonintegrable systems. A measure of this departure is given by the Dyson-Mehta $\Delta_3$ statistic \cite{DysonMehta1963}:
\begin{align}
\label{eq:Delta3}
\Delta_3(L;\mce)\equiv\frac{1}{L}\mathop{\rm Min}_{A,B}\int_\mce^{\mce+L}[N(E)-AE-B]^2dE\,.
\end{align}
Here $N(E)$ is the number of levels with a normalized energy less than $E$. This is a staircase-like function with an approximate slope of unity. $A$ and $B$ are the constants that give a best straight line fit to $N(E)$ in the interval $\mce\leq{}E<\mce+L$. One can define $\bar\Delta_3(L)\equiv\langle\Delta_3(L;\mce)\rangle_\mce$ by averaging over various windows each of length $L$.

The harmonic oscillator gives the least possible value of $\Delta_3=1/12$. For a random spectrum, with a Poisson spacing, $\Delta_3=L/15$, independent of $\mce$. For a GOE, $\bar\Delta_3(L)=(\ln{L}-0.0687)/\pi^2$. Plots of $\bar\Delta_3(L)$ for these spectra are shown along with our result in Figure \ref{fig:rigidity}.

\subsection{Asymptotically integrable systems}\label{sec:ais}
The systems that we encounter in holography are classically chaotic, but only in a certain regime of parameters. For our example of the $AdS$ soliton, the energy $E$ acts as a parameter that dials the transition to chaos \cite{Basu:2011dg}. For large values of energy, the system becomes momentum dominated and cares very little about the details in the potential. We thus expect this system to be approximately integrable for those parameter values. 

This behavior should also be reflected in the quantum spectrum. We should expect to see features of quantum chaos only for certain values of parameters of the system. However if quantum chaos is defined from the distribution of level spacing (\ref{eq:GOE}), then one needs to calculate the differences between a large number of adjacent levels to approach a perfect GOE distribution. However our model enters into a momentum dominated integrable regime at higher energies. Hence we should not expect to find an exact (or even sufficiently close) GOE distribution for any value of parameters, small or large. Typically we should get an intermediate behavior between two extreme cases of GOE and Poisson distribution.\footnote{It is unclear even theoretically what exact level spacing distribution our models should show asymptotically.} We expect this generic behavior of the spectrum to hold in the full theory, as asymptotic integrability is a property of the full quantum theory without any minisuperspace quantization.

\section{Numerical solution and results}\label{sec:results}

In this section we obtain the spectrum for the $AdS$ soliton within the framework of minisuperspace quantization set up in Section~\ref{sec:minisup}. We need to obtain the eigenvalues $E^2$ of (\ref{eq:adssol-effham-yR}). We obtain the eigenvalues numerically using pseudospectral method. For the purpose of numerics, we restrict ourselves to $d=4$ and use $\alpha=1$. We cut the problem down to a finite domain $y_{\rm min}\!<\!y\!<\!y_{\rm max}$, $R_{\rm min}\!<\!R\!<\!R_{\rm max}$ and discretize it on a $N\!\times\!N$ Tchebychef grid \cite{Trefethen}. We impose hard-wall Dirichlet boundary conditions on the boundary of the domain -- therefore it is important to make sure that the true eigenfunction falls off to nearly zero there. This can be ensured by restricting ourselves to eigenvalues $E^2$ much smaller than the value of the potential on the boundary.

\begin{figure}[h!]
\centering
\subfigure[3D perspective plot.]{
\includegraphics[scale=0.8]{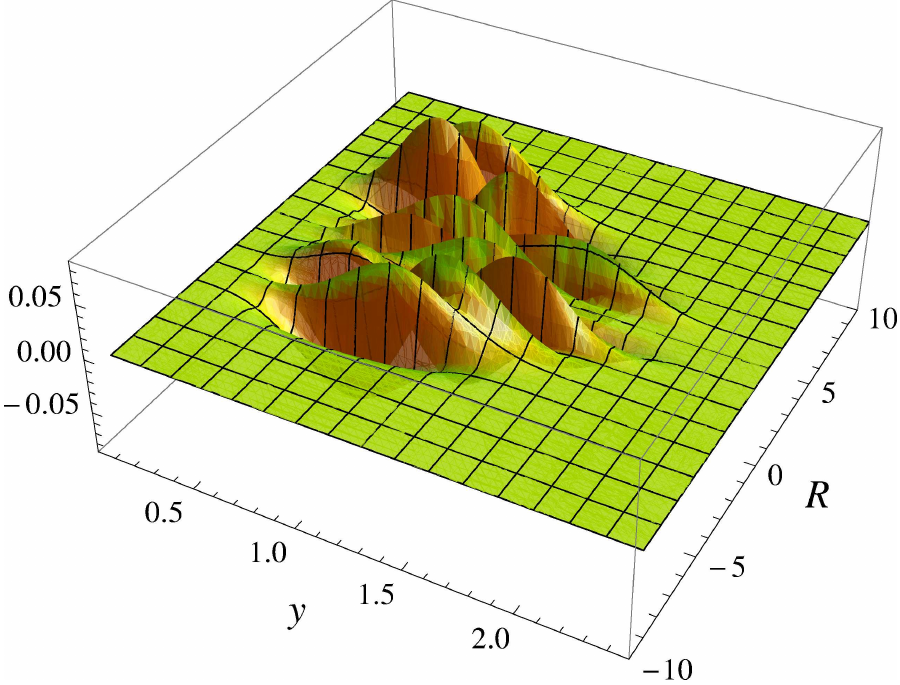}
\label{fig:3D}
}
\subfigure[Contour plot.]{
\includegraphics[scale=0.65]{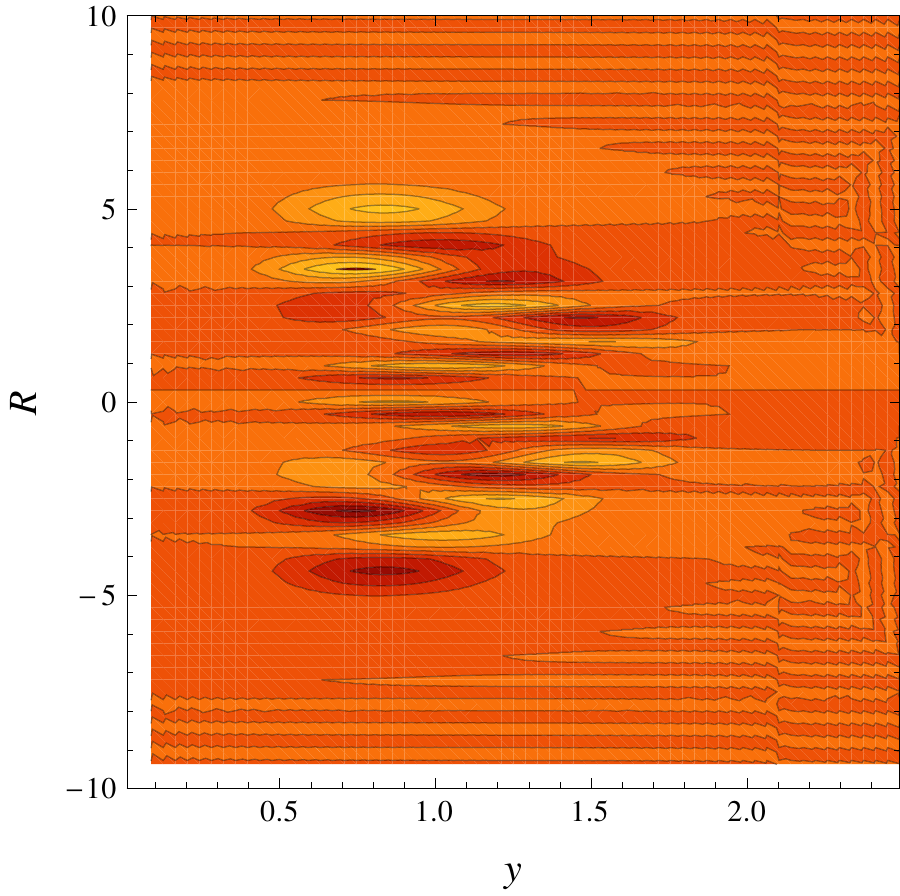}
\label{fig:CONT}
}
\caption{Eigenfunction for the $42^{\rm nd}$ energy level for $k=4.0$ with $E^2\approx98.4$.}
\label{fig:eigenfunction}
\end{figure}

It suffices our purpose to choose $R_{\rm max}\!=10=\!-R_{\rm min}$, $y_{\rm min}\!=0.1$, $y_{\rm max}\!=2.5$. We use $N=64$. From the plots of the potential [Figure~\ref{fig:pot}], we see that $\tilde{V}(y,R)\gtrsim200$ on the boundary. We look at eigenvalues with $E^2\!\lesssim200$ for $4\leq{k}\leq9$. A typical eigenfunction is shown in Figure~\ref{fig:eigenfunction}.

\begin{figure}
\centering
\subfigure[Nearly Wigner GOE for small $E$.]{
\includegraphics[scale=0.38]{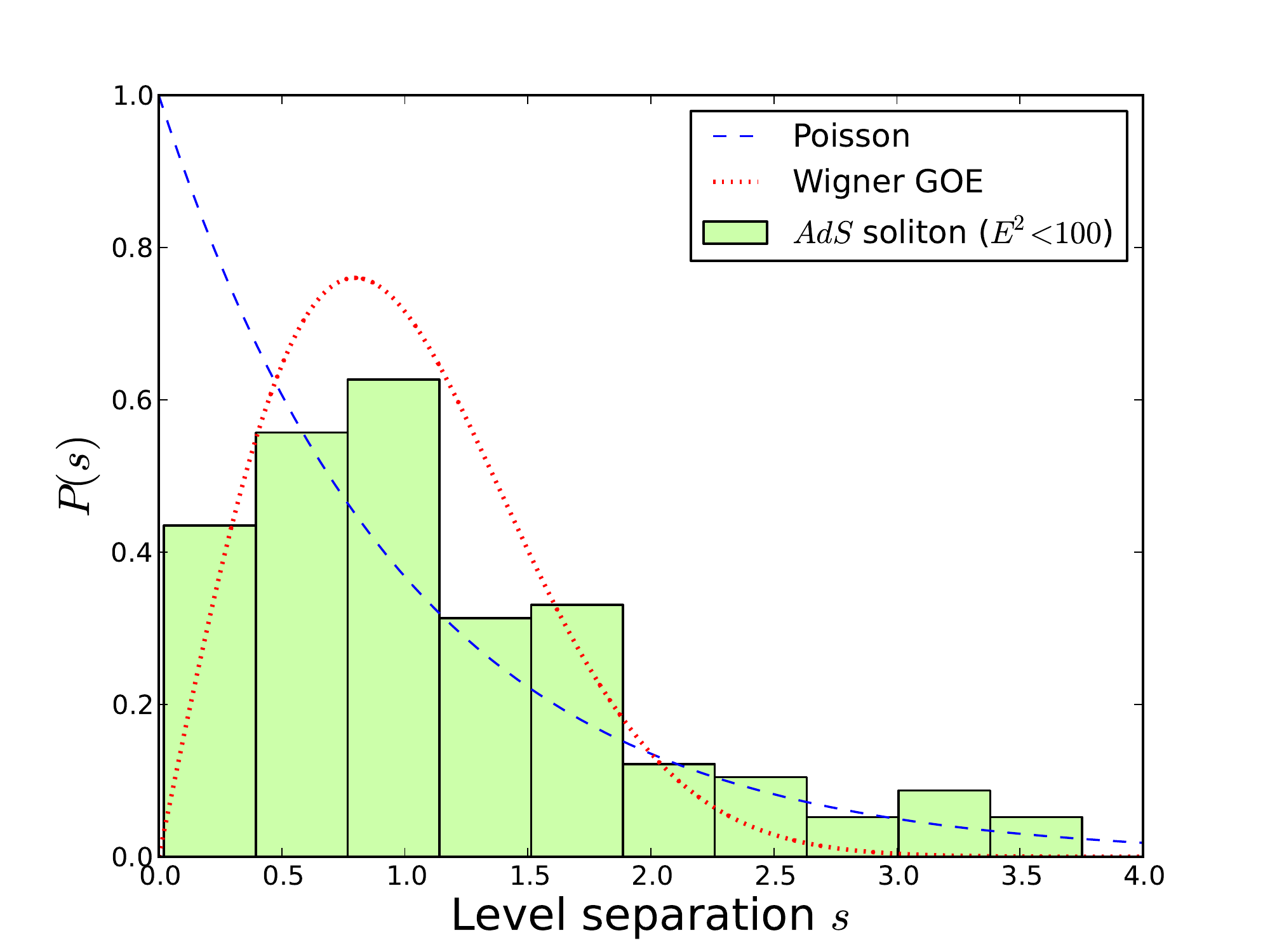}
\label{fig:AdSsolQC}
}
\subfigure[Nearly Poisson for large $E$.]{
\includegraphics[scale=0.38]{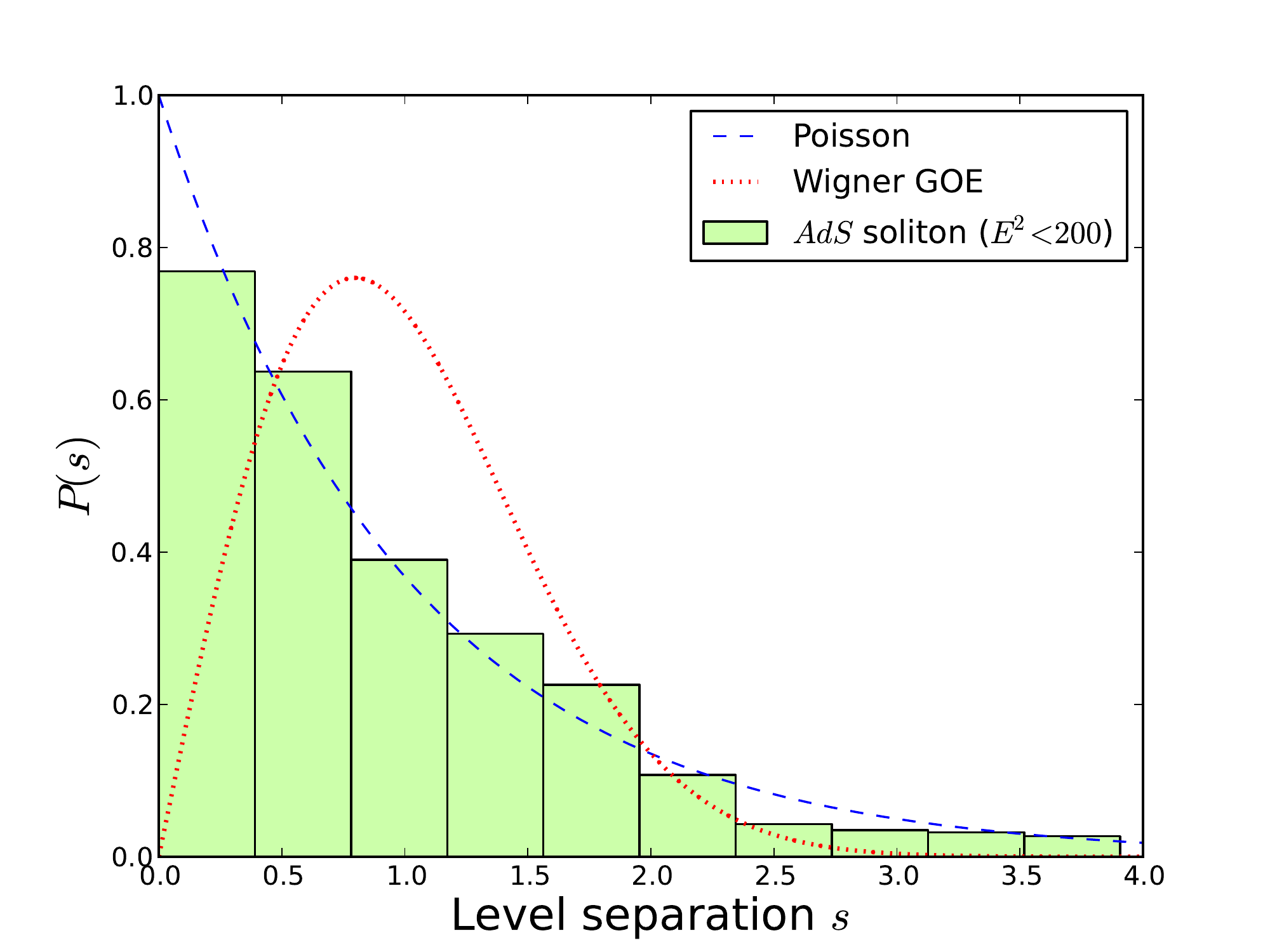}
\label{fig:AdSsolNOQC}
}
\caption{The distribution of level-spacings in the $AdS$ soliton. We have chosen $4\!\leq\!k\!\leq\!9$ and have a total of $167$ differences for $E^2<100$ in the left panel and $994$ differences for $E^2<200$ in the right panel. For  lower energies we see level repulsion and a distribution close to Wigner GOE. For higher energies, this feature is lost and we recover a Poisson-like distribution for uncorrelated levels.
}
\label{fig:soliton-qc}
\end{figure}

After obtaining the energies from the eigenvalues, for each value of $k$, we normalize them, find the difference between the nearest members and plot a histogram of the level spacing distribution. Our results are shown in Figure~\ref{fig:soliton-qc}. Restricting to small values of energy, $E^2<100$, we obtain a distribution similar to the Wigner GOE with a clear signature of level repulsion, indicative of quantum chaos. However, going up to higher energies $E^2<200$, we get back a histogram agreeing with the Poisson distribution of spacing between random uncorrelated levels. This points to the fact that our system is asymptotically integrable and shows features of chaos only at certain intermediate values of energy.

\begin{figure}
\centering
\includegraphics[scale=0.50]{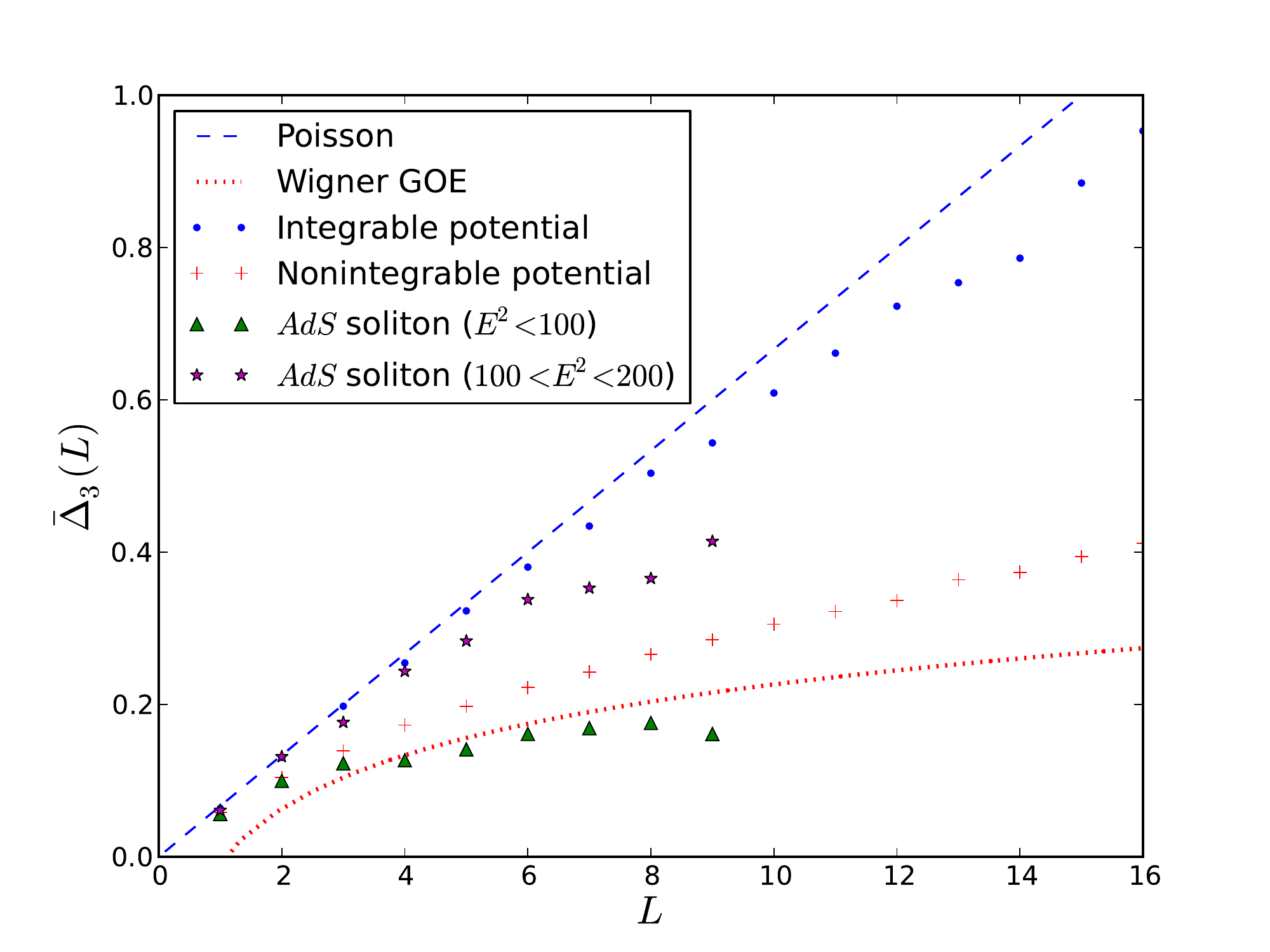}
\caption{Spectral rigidity and Dyson-Mehta $\Delta_3$ statistic. Dashed blue and dotted red curves are for Poisson and Wigner GOE distributions respectively. We see an agreement with $\bar\Delta_3$ for the integrable and nonintegrable potentials of Figure~\ref{fig:qc_box}. For the $AdS$ soliton, we obtain an agreement with Wigner GOE for lower energies (green triangles, $E^2\!<\!100$) and an approximate agreement with Poisson for higher energies (magenta stars, $100\!<\!E^2\!<\!200$).}
\label{fig:rigidity}
\end{figure}

As a further test, we calculate the $\Delta_3$ statistic characterizing spectral rigidity defined in (\ref{eq:Delta3}). This is shown in Figure~\ref{fig:rigidity}. For smaller energies $E^2<100$, we see smaller values $\bar\Delta_3(L)$, along the curve for quantum chaotic systems. For larger values of energy $100<E^2<200$, we see larger values of $\bar\Delta_3$, close to the $L/15$ straight line for random levels, characterizing a large deviation from equal spacing seen in integrable systems.

\section*{Acknowledgements}
This work is partially supported by National Science Foundation grants PHY-0970069, PHY-0855614 and PHY-1214341. We would like to thank Leopoldo Pando Zayas and Diptarka Das for discussion and collaboration during the initial stages of the work, Dori Reichmann for collaboration, discussion and especially for bringing our attention to the spectral methods we use here, Al Shapere and Sumit Das for valuable discussion, and Satya Majumdar for some clarifications during the preparation of the final draft. A.G. would like to thank Shailesh Lal for a lot of valuable discussion in the later stages and Abhishake Sadhukhan for a very stimulating conversation on the day of the Indian festival of Holi nearing the completion of this work.

\bibliographystyle{JHEP}
\bibliography{otherrefs,inspirerefs}

\end{document}